\documentclass[english,preprint]{aastex}

\shorttitle{}
\shortauthors{}

\begin{document}

\title{The Removal of Artificially Generated Polarization in SHARP Maps}

\author{Michael Attard, Martin Houde}

\affil{Department of Physics and Astronomy, The University of Western Ontario,
London, Ontario, Canada N6A 3K7}

\email{mattard@uwo.ca}

\author{Giles Novak}

\affil{Department of Physics and Astronomy, Northwestern University, Evanston,
Illinois, USA, 60208}

\email{g-novak@northwestern.edu}

\and{}

\author{John E. Vaillancourt}

\affil{Department of Physics and Astronomy, California Institute of Technology, Pasadena, California, USA, 91125}

\email{johnv@submm.caltech.edu}

\begin{abstract}
We characterize the problem of artificial polarization for the Submillimeter
High Angular Resolution Polarimeter (SHARP) through the use of simulated
data and observations made at the Caltech Submillimeter Observatory
(CSO). These erroneous, artificial polarization signals are introduced
into the data through misalignments in the bolometer sub-arrays plus
pointing drifts present during the data-taking procedure. An algorithm
is outlined here to address this problem and correct for it, provided
that one can measure the degree of the sub-array misalignments and
telescope pointing drifts. Tests involving simulated sources of Gaussian
intensity profile indicate that the level of introduced artificial
polarization is highly dependent upon the angular size of the source.
Despite this, the correction algorithm is effective at removing up
to 60\% of the artificial polarization during these tests. The analysis
of Jupiter data taken in January 2006 and February 2007 indicates
a mean polarization of $1.44\%\pm0.04\%$ and $0.95\%\pm0.09\%$, respectively. The application of the correction algorithm
yields mean reductions in the polarization of approximately 0.15\%
and 0.03\% for the 2006 and 2007 data sets, respectively. 
\end{abstract}


\section{Introduction}

Submillimeter polarimetry provides a means to investigate the morphology
of interstellar magnetic fields that are highly embedded in dusty
clouds. Such an investigative tool is extremely useful for the study
of astrophysical phenomena in which magnetic fields are suspected
to play a significant role. Such areas of interest include: star formation
\citep{Shu 1987,Hildebrand 1984}, circumstellar disks and jets \citep{Davis 2000},
filamentary structure in molecular clouds \citep{Fiege and Pudritz 2000},
and galactic-scale field morphology \citep{Greaves and Holland 2002}.
In the particular case of low mass star formation, the current leading
model places great emphasis upon the presence of embedded magnetic
fields to regulate the entire process \citep{Mouschovias 2001}. Hence
any further understanding of these magnetic fields may yield a clearer
understanding of the origins of \textquotedblleft{}solar-like\textquotedblright{}
stellar/planetary systems.

Current work in this field is being carried out at the Caltech Submillimeter
Observatory (CSO) using the Submillimeter High Angular Resolution
Polarimeter (SHARP). SHARP is a fore-optics module designed to be
used in conjunction with the SHARC-II camera to form a highly sensitive,
dual-wavelength (350-$\mu\mathrm{m}$ and 450-$\mu\mathrm{m}$), polarimeter
\citep{Novak 2004,Li 2006}. SHARC-II employs a 12 $\times$ 32 pixel bolometer
array that is optically \textquotedblleft{}split\textquotedblright{}
by SHARP into three zones: two 12 $\times$ 12 pixel regions that record
orthogonal states of linear polarization (which are labeled \textquotedblleft{}H\textquotedblright{}
and \textquotedblleft{}V\textquotedblright{} for horizontal and vertical,
respectively), and a 12 $\times$ 8 pixel central zone that is not used with
SHARP. The horizontal and vertical components are combined during
data reduction to yield the $I$, $Q$, and $U$ Stokes parameters.

The simultaneous measurement of the H and V polarization components
allows for the effective removal of the sky background signal \citep{Hildebrand 2000}.
However, it does not negate the possibility of erroneous polarization
signal generation. The combination of misalignments between the two
sub-arrays (i.e., H and V) and pointing drifts during the observation
cycle, can result in the generation of artificial polarization. The
generation of these erroneous signals may place limitations on the
sensitivity of SHARP and thus could reduce data-gathering efficiency.
This would hurt efforts to rapidly survey large extended objects,
such as giant molecular clouds (GMC\textquoteright{}s), where many
observations would be required to properly survey the source and thus
a high data-taking efficiency is required. 

A correction algorithm has been designed in an attempt to model and
correct for this problem in the SHARP data reduction pipeline. This
paper will go over in detail the problem of artificial polarization
in dual-array polarimeters and the algorithm by which a correction
is attempted, with simulated and planetary data being used to test
the proposed method. Section \ref{sec:2} will describe the means
by which artificial polarization is generated in a dual-array polarimeter.
Section \ref{sec:3} will describe the algorithm employed to treat
this problem. Section \ref{sec:4} of this paper will discuss the
magnitude of the problem and cover the results obtained thus far from
the testing of simulated and planetary data. Section \ref{sec:5}
covers the concluding remarks.

\section{\label{sec:2}Artificial Polarization}

Figure 1 illustrates how the SHARC-II array is segmented into three
regions; the aforementioned \textquotedblleft{}H\textquotedblright{}
and \textquotedblleft{}V\textquotedblright{} sub-arrays for the horizontal
and vertical polarization components, respectively, and a central
unused zone. Note that horizontal and vertical are defined with respect
to the long axis of the bolometer array, which in the case of Figure
1 is the axis parallel to the horizontal of the image. Consider the
radiation beam that is incident to the H and V sub-arrays. This radiation
beam originates from a single patch of sky that is subsequently \textquotedblleft{}split\textquotedblright{}
into two components; a horizontally polarized component and a vertically
polarized component \citep{Novak 2004}. As the name sake would suggest,
the optical path of SHARP is designed such that the horizontally polarized
component is incident on the \textquotedblleft{}H\textquotedblright{}
sub-array while the vertically polarized component is incident on
the \textquotedblleft{}V\textquotedblright{} sub-array. In this way,
both arrays image the same patch of sky. Also consider two position
vectors, $\mathbf{x_{\mathrm{H}}}$ and $\mathbf{x_{\mathrm{V}}}$,
that we will use to map the H and V sub-arrays, respectively. Note
that each of these vectors has an independent origin (in their respective
sub-arrays). Once the incident radiation is absorbed by the bolometers,
we can express the resultant flux as being a function of the position
vectors for both the H and V sub-arrays, $f_{\mathrm{H}}(\mathbf{x_{\mathrm{H}}})$
and $f_{\mathrm{V}}(\mathbf{x_{\mathrm{V}}})$ , respectively. 

Now the two position vectors will be related through:

\begin{equation}
\mathbf{x_{\mathrm{V}}}=\mathbf{d}+\mathbf{R}\mathbf{S\mathbf{x_{\mathrm{H}}}},\end{equation}

\noindent where the quantities $\mathbf{d}$, $\mathbf{R}$, and $\mathbf{S}$
are the V array translational displacement, rotation, and stretch
matrices relative to the H array, which is taken as a reference%
\footnote{Lower case bold letters represent vector quantities, while upper case
bold letters represent matrices. This convention will be held throughout
the paper.%
}. Note that the stretch matrix describes a magnification/minification of the image on the sub-array. In an ideal setting we would have: $\mathbf{d}=\mathbf{0}$, $\mathbf{R}=\mathbf{1}$,
and $\mathbf{S}=\mathbf{1}$ where $\mathbf{0}$ is a \textquotedblleft{}zero\textquotedblright{}
vector and $\mathbf{1}$ is the identity matrix. This would imply
no array misalignments and $\mathbf{x_{\mathrm{H}}}=\mathbf{x_{\mathrm{V}}}=\mathbf{x}$.
As will be explained later, in this case any measured polarization
would result from either: 1) the detection of a polarized source or,
2) instrumental polarization. In reality, however, small misalignments
between the arrays are present and complicate the interpretation of
the polarization data.

During one cycles-worth of observations, measurements of $f_{\mathrm{H}}(\mathbf{x_{\mathrm{H}}})$
and $f_{\mathrm{V}}(\mathbf{x_{\mathrm{V}}})$ are made for each of
the four half wave plate (HWP) angular positions: $\theta=0^{\circ},\,22.5^{\circ},\,45^{\circ},\,$ and $67.5^{\circ}$.
The effect of rotating the HWP is to rotate the polarization of the
incoming signal by $2\theta$ \citep{Hildebrand 2000}. This enables
the flux of the signal to be measured with its incident state of linear
polarization rotated by angles of $\;0^{\circ},\,45^{\circ},\,90^{\circ},\,135^{\circ}$
and thus allows for the calculation of the Stokes parameters. Note
that only the linear polarization can be determined with this methodology,
as measurements of circular polarization would require the use of
a quarter wave plate. What is obtained in the end are eight flux maps,
four H array maps and four V array maps, that can then be processed
to generate images of the Stokes parameters $I$, $Q$, and $U$.
These parameters are given by:

\begin{eqnarray}
I & = & \frac{1}{4}\{f_{\mathrm{H}}(\mathbf{x_{\mathrm{H}}^{\mathrm{0^{\circ}}}})+f_{\mathrm{V}}(\mathbf{x_{\mathrm{V}}^{\mathrm{0^{\circ}}}})+f_{\mathrm{H}}(\mathbf{x_{\mathrm{H}}^{\mathrm{22.5^{\circ}}}})+f_{\mathrm{V}}(\mathbf{x_{\mathrm{V}}^{\mathrm{22.5^{\circ}}}})\} \nonumber \\
& & +\frac{1}{4}\{f_{\mathrm{H}}(\mathbf{x_{\mathrm{H}}^{\mathrm{45^{\circ}}}})+f_{\mathrm{V}}(\mathbf{x_{\mathrm{V}}^{\mathrm{45^{\circ}}}})+f_{\mathrm{H}}(\mathbf{x_{\mathrm{H}}^{\mathrm{67.5^{\circ}}}})+f_{\mathrm{V}}(\mathbf{x_{\mathrm{V}}^{\mathrm{67.5^{\circ}}}})\}\\
Q & = & \frac{1}{2}\{[f_{\mathrm{H}}(\mathbf{x_{\mathrm{H}}^{\mathrm{0^{\circ}}}})-f_{\mathrm{V}}(\mathbf{x_{\mathrm{V}}^{\mathrm{0^{\circ}}}})]-[f_{\mathrm{H}}(\mathbf{x_{\mathrm{H}}^{\mathrm{45^{\circ}}}})-f_{\mathrm{V}}(\mathbf{x_{\mathrm{V}}^{\mathrm{45^{\circ}}}})]\}\\
U & = & -\frac{1}{2}\{[f_{\mathrm{H}}(\mathbf{x_{\mathrm{H}}^{\mathrm{22.5^{\circ}}}})-f_{\mathrm{V}}(\mathbf{x_{\mathrm{V}}^{\mathrm{22.5^{\circ}}}})]-[f_{\mathrm{H}}(\mathbf{x_{\mathrm{H}}^{\mathrm{67.5^{\circ}}}})-f_{\mathrm{V}}(\mathbf{x_{\mathrm{V}}^{\mathrm{67.5^{\circ}}}})]\},\end{eqnarray}

\noindent where $\mathbf{x_{\mathrm{i}}^{\theta}}=\mathbf{\mathrm{\mathbf{x}}_{\mathrm{i}}}+\mathbf{p^{\theta}}$
with $\theta=22.5^{\circ},\,45^{\circ},\,$ and $67.5^{\circ}$ for
the HWP angles, and $\mathbf{\mathrm{i\,}}$= \{H,V\}. The $\mathbf{p^{\theta}}$
vectors represent the mean telescope pointing drift at the $\theta$
HWP angle with respect to the reference $\mathbf{p^{\mathrm{0}}}=\mathbf{0}$
. This implies that we must treat the flux as being a function of
$\theta$, as well as position on the sub-arrays; this is included
in the notation of equations (2), (3), and (4). Ideally, observations
would not suffer from pointing errors and thus $\mathbf{p}^{\theta}=\mathbf{0}$
, regardless of the HWP angle. However, in reality the pointing will
drift by some amount over the course of the cycle. Note that the nature of this pointing drift is random, systematic shifts in the telescope pointing over the course of one modulation cycle.

We are now in a position to study the root causes of artificial polarization.
For the purpose of this illustration let us assume we are dealing
with an unpolarized source. If misalignments exist between the H and
V arrays such that the pixel space coordinates between the two sub-arrays
are related by equation (1), then for any position on the array the
quantity $T$:

\begin{equation}
T(\mathbf{x}_{\mathrm{H}}^{\theta},\mathbf{x_{\mathrm{\mathrm{V}}}^{\theta}})\equiv f_{\mathrm{H}}(\mathbf{x}_{\mathrm{H}}^{\theta})-f_{\mathrm{\mathrm{V}}}(\mathbf{x}_{\mathrm{V}}^{\theta})\end{equation}

\noindent will be non-zero. However, since each expression for $Q$
and $U$ contains the difference (denoted as $M$) between such terms:

\begin{equation}
M\equiv T(\mathbf{x}_{\mathrm{H}}^{\theta},\mathbf{x_{\mathrm{\mathrm{V}}}^{\theta}})-T(\mathbf{x}_{\mathrm{H}}^{\theta+45^{\circ}},\mathbf{x_{\mathrm{\mathrm{V}}}^{\theta+\mathrm{45^{\circ}}}}),\end{equation}

\noindent then these non-zero values will cancel each other out provided
there is no pointing drift, $\mathbf{p^{\theta}}$, between HWP positions.
This is due to our assumption that the signal is unpolarized. 

If a pointing drift is present, each $f_{\mathrm{i}}(\mathbf{x}_{\mathrm{i}}^{\theta})$
term (where $\mathrm{i}$ = \{H,V\}) in equations (2), (3), and (4)
would represent the flux of the source offset with respect to the
reference position at $\theta=0^{\circ}$. The presence of these offsets
between HWP positions could prevent the cancellation in equation (6)
of the non-zero difference terms in equation (5) originating from
array misalignments. Only if the source flux $f_{\mathrm{i}}(\mathbf{x}_{\mathrm{i}}^{\theta})$
has a linear gradient over the image (or none at all, in which case
we would be dealing with a flat field) will the combination of array
misalignments plus pointing drifts cause no artificial polarization.
This is because $M=0$ for sources with linear gradients regardless
of any pointing drifts or array misalignments that may be present
during data collection.

In the most general case however, the source fluxes $f_{\mathrm{i}}(\mathbf{x}_{\mathrm{i}}^{\theta})$
will have non-linear gradients over the array, and pointing drifts
and misalignments will be present. In this case there is nothing to
prevent the Stokes $Q$ and $U$ parameters from acquiring non-zero
values for some positions, even if the instrumental polarization is
fully removed from the data and the source is completely unpolarized.

\section{\label{sec:3}Algorithm for Corrections}

We begin by first assuming that the values for $\mathbf{d}$, $\mathbf{R}$,
$\mathbf{S}$, and $\mathbf{p^{\theta}}$ are known. Section \ref{sub: 4.1}
will briefly discuss how these quantities are actually measured with
SHARP. To remove the artificial polarization from the data, the array
misalignments and pointing drifts that would normally distort the
H and V maps must be corrected. Consider an arbitrary position vector
$\mathbf{a}$ specifying a position on a given source. The goal here
is to setup the corresponding position vectors ($\mathbf{a_{\mathrm{H}}}$
and $\mathbf{a_{\mathrm{V}}}$) for the sub-arrays. This is illustrated
below in equations (7) and (8):

\begin{eqnarray}
\mathbf{a_{\mathrm{H}}^{\theta}} & = & \mathbf{a-p^{\theta}}\\
\mathbf{a_{\mathrm{V}}^{\theta}} & = & \mathbf{S^{\mathrm{-1}}R^{\mathrm{-1}}(a-d-p^{\theta})},\end{eqnarray}

\noindent where $\mathbf{RR^{\mathrm{-1}}=\mathbf{1}}$ and $\mathbf{SS^{\mathrm{-1}}=\mathbf{1}}$.
Now the flux measured on the two sub arrays at the positions corresponding
to $\mathbf{a}$ can be expressed as:

\begin{eqnarray}
H(\mathbf{a},\theta) & = & f_{\mathrm{H}}(\mathbf{a}_{\mathrm{H}}^{\theta})\\
V(\mathbf{a},\theta) & = & f_{\mathrm{V}}(\mathbf{a_{\mathrm{V}}^{\theta}}).\end{eqnarray}

The fluxes $H(\mathbf{a},\theta)$ and $V(\mathbf{a},\theta)$ are
now used to compute $Q$ and $U$ maps that are free of artificial
polarization%
\footnote{The actual algorithm currently used for SHARP data analysis does not
exactly follow the methodology outlined in Section \ref{sec:3}, but
the method presented here is mathematically equivalent and simpler
to follow. %
} :

\begin{eqnarray}
I(\mathbf{a}) & = & \frac{1}{4}\{H(\mathbf{a},0^{\circ})+V(\mathbf{a},0^{\circ})+H(\mathbf{a},22.5^{\circ})+V(\mathbf{a},22.5^{\circ})\} \nonumber \\
& & +\frac{1}{4}\{H(\mathbf{a},45^{\circ})+V(\mathbf{a},45^{\circ})+H(\mathbf{a},67.5^{\circ})+V(\mathbf{a},67.5^{\circ})\}\label{eq:I}\\
Q(\mathbf{a}) & = & \frac{1}{2}\{[H(\mathbf{a},0^{\circ})-V(\mathbf{a},0^{\circ})]-[H(\mathbf{a},45^{\circ})-V(\mathbf{a},45^{\circ})]\}\label{eq:Q}\\
U\left(\mathbf{a}\right) & = & -\frac{1}{2}\{[H(\mathbf{a},22.5^{\circ})-V(\mathbf{a},22.5^{\circ})]-[H(\mathbf{a},67.5^{\circ})-V(\mathbf{a},67.5^{\circ})]\}.\label{eq:U}\end{eqnarray}

\section{\label{sec:4}Results}

This section is subdivided into three portions; a brief description
of the observed hardware misalignments and pointing drifts, the degree
to which artificial polarization affects polarimetry data, and results
from simulated and planetary data.

\subsection{\label{sub: 4.1}Measured Hardware Misalignments and Pointing Drifts}

The stretches, rotation angles, and translations of the H and V SHARC-II
bolometer sub-arrays can be measured by placing an opaque plastic
disk in the optical path before SHARP with five pinholes drilled through
it. The pinholes are arranged in a \textquotedblleft{}cross-pattern\textquotedblright{}
with the central hole approximately aligned with the middle of the
sub-arrays and the remaining four holes placed equidistantly from
this central position. Data taken with this disk in place and a uniform
background source (e.g., a cold load) can be analyzed to yield the
hardware misalignments. For observing runs where no alignment data
is taken with the opaque disk, the translations $\mathbf{d}$ can
still be measured by comparing the centroid positions of images on
the sky (e.g., for Jupiter observations) in the H and V sub-arrays.
Typical values include a negligible stretch and a relative rotation
of $\thickapprox2^{\circ}-3^{\circ}$. During the two periods in which
the planetary data to be discussed later were taken, the translational
misalignments were measured to be:

\begin{eqnarray}
(d_{x}\pm\delta d_{x},d_{y}\pm\delta d_{y}) & = & (-0.45\pm0.07,-0.11\pm0.04)\,\mathrm{pixels}\,\mathrm{[January\,2006]}\label{eq:dx06}\\
(d_{x}\pm\delta d_{x},d_{y}\pm\delta d_{y}) & = & (-0.02\pm0.12,-0.41\pm0.05)\mathrm{\, pixels}\,\mathrm{[February\,2007]},\label{eq:dx07}\end{eqnarray}

\noindent where $d_{x}$ and $d_{y}$ are directed along the horizontal
and vertical axes of the bolometer, respectively. Note that negative
signs imply the V sub-array is shifted to the right, or down, of the
H sub-array for an observer looking along the SHARP optical path towards
the bolometer array. The net maximum translation is calculated to
be approximately $\thickapprox0.47$ and $\thickapprox0.43$ pixels
for the 2006 and 2007 observing runs, respectively.  This net maximum translation is calculated by adding in quadrature the horizontal and vertical means and standard deviations.   

The pointing drifts are measured via a correlation program that analyzes
the intensity maps for a given source at each of the four HWP positions
sequenced through during a cycle. The intensity map at $\theta=0^{\circ}$
is taken as the reference for this analysis. The results vary with
each observing run and weather conditions. However, the mean pointing
drifts measured in January 2006 and February 2007 are: \begin{eqnarray}
(p_{x}\pm\delta p_{x},p_{y}\pm\delta p_{y}) & = & (0.03\pm0.20,0.01\pm0.10)\mathrm{\, pixels}\,\mathrm{[January\,2006]}\label{eq:px06}\\
(p_{x}\pm\delta p_{x},p_{y}\pm\delta p_{y}) & = & (0.01\pm0.12,-0.02\pm0.10)\,\mathrm{pixels}\,\mathrm{[February\,2007]},\label{eq:px07}\end{eqnarray}

\noindent where $p_{x}$ and $p_{y}$ are directed along the horizontal
and vertical axes of the bolometer, respectively. It is apparent that
there is a considerable spread about the mean drift magnitude. The
net maximum pointing drift is thus calculated to be $\thickapprox0.23$
and $\thickapprox0.16$ pixels per HWP position for the 2006 and 2007
observing runs, respectively. This net maximum pointing drift is calculated by adding in quadrature the horizontal and vertical means and standard deviations. Each HWP position requires approximately 1.81 minutes of integration time when using SHARP.

\subsection{\label{sub:4.2}A Measure of the Artificial Polarization Problem}

Simulated data are generated as Gaussian sources with various elliptical
aspect ratios. In addition to this, artificial hardware misalignments
and pointing drifts can be introduced into the data. For the purpose
of this discussion three unpolarized simulated sources were generated:
a 9\arcsec circular, a 20\arcsec circular, and a 10\arcsec $\times$ 15\arcsec
elliptical Gaussians (note that one SHARP pixel is approximately 4.6\arcsec
$\times$ 4.6\arcsec). These dimensions refer to the Full-Width-Half-Magnitudes
(FWHM) of the source. These data were generated with no bad pixels
in the array and no noise. The sources were subjected to a range of
hardware misalignments and pointing drifts. The results are presented
in Figure 2.

It should be noted that in our simulation software the pointing drifts are introduced into the data cycle by selecting a magnitude $m$ and direction represented by a unit vector $\mathbf{e}_{i}$.  Then for each HWP position ($\theta=0^{\circ},\,22.5^{\circ},\,45^{\circ},\,67.5^{\circ}$) the following drifts were introduced into the data: $0$, $m\mathbf{e}_{i}$, $2m\mathbf{e}_{i}$, and $-m\mathbf{e}_{i}$, respectively.  This is hardly a random pointing drift; in fact each displacement lies on a line defined by the unit vector $\mathbf{e}_{i}$. Therefore it is easy to conclude that our modeling of the pointing drift has limitations when compared with the random, systematic drifts that are present in real data. 

One notices immediately the varying magnitude of the artificial polarization
illustrated over the three plots. The 9\arcsec circular Gaussian
generates roughly 8\% of artificial polarization for a 0.5 pixel
translation and a pointing drift of one SHARP pixel (i.e., 4.6\arcsec)
per HWP position, while the 20\arcsec circular Gaussian generates
only about 0.4\% for the same misalignments and drifts. This trend
is directly related to the broadness of the source; a more compact
source will have a larger intensity gradient across its profile and
as such a large polarization is induced due to the abrupt change in
intensity with position. To understand this effect better it is instructive
to compare the actual maps of the Stokes parameters $I$, $Q$, and
$U$ for these simulated sources. These are presented in Figure 3 for the case of a 4.6 arcseconds per HWP position pointing drift (in the horizontal direction) and a 0.5 pixel translation between the H and V sub-arrays (in the vertical direction). The alternating light-dark pattern seen in the $Q$ and $U$ images results from the fact that the pointing drift and array translation are in orthogonal directions, and from the shape of the source itself. The $Q$ and $U$ images look identical as the simulated source is unpolarized. As a result, equations 3 \& 4 will have no dependance on the HWP angle and are thus mathematically equivalent. One should note that the maps of $Q$ and $U$ illustrated in this figure would be flat, uniform fields if no artificial linear polarization were
detected from any of the sources. The results are contrary to this however, with structure being apparent in the $Q$ and $U$ maps for
each of the simulated sources.

Referring to Figure 2c, one can see that for typical values of array
misalignment observed with SHARP the effect of rotations will play
a secondary role to that of translations. Stretches were not tested
as measurements with SHARP indicate that they are negligible.

\subsection{\label{sub:4.3}Simulated and Planetary Data Results}

\subsubsection{\label{sub:4.3.1}Corrections for Simulated Data with no Noise and
no Bad Pixels}

Simulated data provide the first test for the effectiveness of the
algorithm outlined in Section \ref{sec:3}. These provide ideal cases,
as the hardware misalignments are known precisely. In addition, the
correlation routine used to measure the pointing drifts can be tested
under controlled conditions. It is typically found that the pointing
can be measured to an accuracy of $\pm0.01$ pixels with no noise
present in the signal and no bad pixels in the array. We now look
again to the three simulated sources discussed in the previous subsection
to see how effectively the artificial polarization can be removed.
The results are illustrated in Figure 4. 

It is clear from this figure that a significant reduction in the polarization
level is achieved after the corrections are made. The most significant
reduction is evident in the elliptical and largest circular cases,
where the polarization is truncated by approximately 50\% -60\%. The small
circular case shows an improvement in the polarization level of approximately 40\%. Again a significant dependence upon
source size is observed, with larger extended sources showing both
lower induced polarization levels and a lower residual signal level
after correction.

\subsubsection{\label{sub:4.3.2}Corrections for Simulated Data with Noise and Bad
Pixels}

In order to generate simulated data that more accurately reflects
real data, it is desired to measure the performance of the correction
algorithm with simulated data that include noise and bad pixels.
To this end the analysis of the large 20\arcsec circular Gaussian
was redone as it most closely resembles the profile of Jupiter, a
source that will be discussed later in this section. Forty-five bad
pixels were introduced into the simulation; compared with thirty-seven
bad pixels identified in the sub-arrays from data obtained in February
2007. Sufficient noise was introduced to allow for a signal-to-noise
ratio (SNR) of $\thickapprox4.3$ in the data. By introducing bad
pixels and noise it is found that the pointing can be measured to
an accuracy of $\pm0.05$ pixels. The results are presented in Figure
5.

A comparison of Figure 5 with Figure 4b shows that for the SNR considered
here, artificial polarization can be effectively corrected for pointing
drifts approximately greater than 2\arcsec per HWP position and sub-array
misalignments approximately greater than 0.1 pixel. In cases of higher
SNRs, the effects of the noise level will be reduced. In this case
the noise introduces a background polarization level in Figure 5 of
around 0.32\% that washes out all but the most prominent artificial
signal. It should be noted here that although the mean value of the
$Q$ and $U$ Stokes parameters induced due to noise is approximately
zero, the polarization percentage ($P=\sqrt{(Q/I)^{2}+(U/I)^{2}}$
) is an unsigned quantity, resulting in the offset. However, the correction
algorithm does appear to be effective at reducing this artificial
signal down to the background level for larger array translations
(the 0.5 pixel curve) and pointing drifts (2.3\arcsec per HWP position
or more). This example illustrates that when looking at real data
later on it will be essential to take note of the magnitude of the
pointing drift and hardware misalignments, as well as the level of
background noise.

\subsubsection{\label{sub:4.3.3}Corrections for Simulated Data with Noise, Bad
Pixels, and Translation Measurement Errors}

Before discussing the results obtained for the Jupiter data, it is
first desired to talk about the effects of inaccuracies in the hardware
misalignment parameters. Until now, the analysis presented here has
assumed a perfectly accurate knowledge of the misalignment between
the two sub-arrays. This is not reflective of reality. To investigate
how sensitive the correction algorithm is to inaccuracies in the hardware
parameters, simulations were again run of the large 20\arcsec circular
Gaussian. Bad pixels and detector noise were again included into the
data. Known inaccuracies in the hardware parameters were then introduced
into the correction algorithm. The results are presented in Figure
6.

As can be seen from the figure, for errors smaller than $\thickapprox0.1$
pixel the analysis shows that the correction algorithm is degraded
by only a small amount. More precisely, looking at pointing drifts
of 2.3\arcsec or larger, the residual polarized signal is increased
by approximately $\Delta P=0.05\%$ relative to the case where the
hardware misalignments is perfectly known (only larger pointing drifts
were included in the error calculation as drifts smaller than 2.3\arcsec
do not appear to generate a significant artificial polarization signal
above the noise level, as indicated in Figure 5). These results indicate
a degradation of approximately 15\% in the correction algorithm when
compared to the \textquotedblleft{}ideal\textquotedblright{} performance
conditions with no measurement errors. For the milder case of a 0.05
pixel error, the residual signal is found to have increased by $\Delta P=0.03\%$
relative to the case with no errors. This implies a 9\% degradation
in the correction algorithm when compared to ideal conditions. As
we shall see, measurement uncertainties on the order of 0.05 to 0.1
pixels will be close to what is obtained with actual planetary data.

\subsubsection{\label{sub:4.3.4}Corrections for Planetary Data}

Two sets of Jupiter data, obtained in January 2006 and February 2007,
were analyzed in the course of this study. The raw (uncorrected) data
shows a mean of the unsigned levels of polarization in the central
8 pixel by 8 pixel portion of the array to be $\thickapprox1.44\%\pm0.04\%$
and $\thickapprox0.95\%\pm0.09\%$ for the January and February data
sets, respectively. The contribution of the polarization due to the
mean RMS noise levels is found to be $\thicksim0.02\%$ for both data
sets, which is a figure small enough to be accounted for within the
scatter of the mean polarization values. 

For the purposes of this preliminary study, only translational sub-array
misalignments were measured and corrected for. The results of simulation
tests presented in Figure 4c appear to indicate that with the hardware
misalignments and pointing drifts mentioned in Section \ref{sub: 4.1},
the artificial polarization will be dominated by the contribution
originating from translation. 

The Jupiter data analysis results are presented below in Figure 7.
Curves are shown for the raw (uncorrected) signal and the residual
signal from the corrected data as a function of cycle number. 

After corrections, a residual polarization of $1.30\%\pm0.03\%$ and
$0.93\%\pm0.09\%$ is calculated for the 2006 and 2007 Jupiter data
sets, respectively. This indicates an overall reduction in the polarization
by $0.15\%\pm0.01\%$ (i.e., on average the artificial polarization was reduced within a range of approximately $0.14\%$ to $0.16\%$) and $0.03\%\pm0.03\%$ (i.e., on average the artificial polarization was reduced within a range of approximately $0\%$ to $0.06\%$), respectively. These values were calculated by taking the difference between each raw datum and the corresponding residual. The mean and standard deviation of these differences can then be computed to yield the aforementioned reduction values. There
is considerable spread in the data, but a net reduction in the polarization
of the data is observed within the error bars. The less impressive
reduction observed for the February 2007 data set may be due to improved
intra-cycle pointing and the elimination of beam distortions with
one of the sub-array\textquoteright{}s that were present during the
January 2006 observing run \citep{Li 2006}. Considering the magnitude
of the translational misalignments and pointing drifts for the planetary
data discussed here (see eqs. {[}\ref{eq:dx06}] to {[}\ref{eq:px07}]),
one would not expect a dramatic reduction in the polarization. In
fact, these results are consistent with the simulations discussed
previously (see figs. 4b and 5). The fact that a net reduction is
observed can be interpreted as a good indicator that the correction
algorithm is effective at removing some of the artificial polarization. 

It should be clarified here that we are not proposing the correction algorithm can compensate for the beam distortions.  Instead, the presence of these distortions would degrade the quality of the January 2006 data and may account for the increased level of polarization in the raw signal. It is hypothesized here that this degraded data might respond better to the application of the correction algorithm, although a detailed description of how this occurs is not known. It is not claimed here that the modeling described in Sections 4.2, 4.3.1, 4.3.2, and 4.3.3 can fully explain the results obtained on Jupiter.  We merely set out to describe the effect of the correction algorithm on real data and compare those results with the modeling that has been done to date.  There are important differences between the simulated sources and Jupiter.  These include: the planets disk does not have a Gaussian profile and the pointing drifts in real data are directed randomly, not in the linear fashion used in our simulations.  

\section{\label{sec:5}Conclusion}

The correction algorithm proposed in Section \ref{sec:3} has been
effectively tested with simulated and planetary data obtained with
the SHARP. Analysis with simulated data indicates a maximum reduction
in the artificial signal by roughly 60\%. Translational misalignments
in the sub-arrays appear to provide the dominant contribution to artificial
polarization in SHARP, with stretches and rotations being either negligible
or only minor contributors. The correction algorithm appears to be
effective at removing artificial polarization signals from simulated
sources even with the introduction of noise, bad pixels, and uncertainties
in the hardware misalignment measurements.

Reductions of $\thickapprox0.15\%$ (January 2006) and $\thickapprox0.03\%$
(February 2007) in the raw polarimetry signal were achieved with the
correction algorithm on Jupiter data. Considering the difference in
pointing drifts measured during the 2006 and 2007 observing runs (see
eqs. {[}\ref{eq:px06}] to {[}\ref{eq:px07}]), these reductions are
consistent with our simulation results. The residual polarization
signals obtained are $1.30\%\pm0.03\%$ and $0.93\%\pm0.09\%$ for
the 2006 and 2007 Jupiter data sets, respectively.

One should note that the reductions achieved with Jupiter data are
roughly equivalent to the magnitude of the instrumentation polarization
(IP) for this instrument. Therefore, the application of our correction 
algorithm presents approximately the same degree of improvement in the data as the removal of the IP. For example, the published mean IP contribution for the previous
 CSO polarimeter, HERTZ,
is $0.22\%$ for the telescope and within the range of $0.23\%-0.38\%$
for the polarimeter (this value varies over the bolometer array \citep{hertzarchive}).
The IP for SHARP is currently estimated to be approximately twice
as large as that measured for HERTZ, and could account for some of
the polarization remaining in the Jupiter data after we applied our
corrections, especially for the February 2007 data. The bulk of the
residual signal in the 2006 data set might be better explained as
a result of the beam distortions that are known to have been present
in the instrument at that time.

\acknowledgements{}

M.A.'s and M.H.'s research is funded through the NSERC Discovery
Grant, Canada Research Chair, Canada Foundation for Innovation, Ontario
Innovation Trust, and Western's Academic Development Fund programs.
G.N. acknowledges support from NSF grants AST 02-43156 and AST 05-05230
to Northwestern University. J.E.V. acknowledges support from NSF grants AST 05-40882 to the California Institute of Technology and AST 05-05124 to the University of Chicago. SHARC II is funded through the NSF grant
AST 05-40882 to the California Institute of Technology. SHARP is also
funded by the NSF award AST-05-05124 to the University of Chicago.

\clearpage

\begin{figure}
\epsscale{0.5}
\plotone{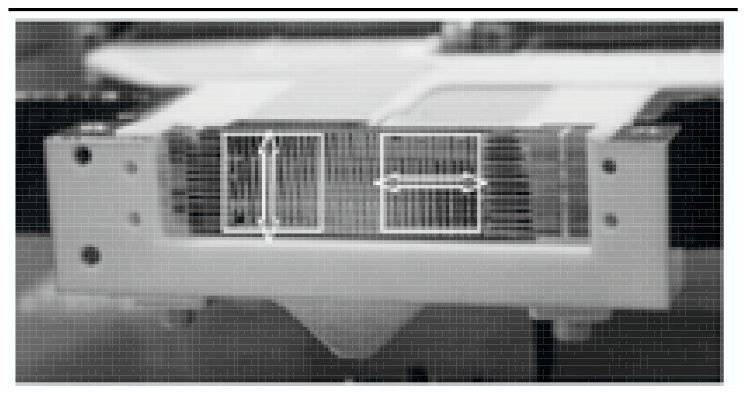}

\caption{The SHARC-II bolometer array. The outlined squares with arrows indicate
the vertical (V) and horizontal (H) sub-arrays. The central region
is a dead zone (from Li et al. 2006). }

\end{figure}

\begin{figure}
\epsscale{0.5}
\plotone{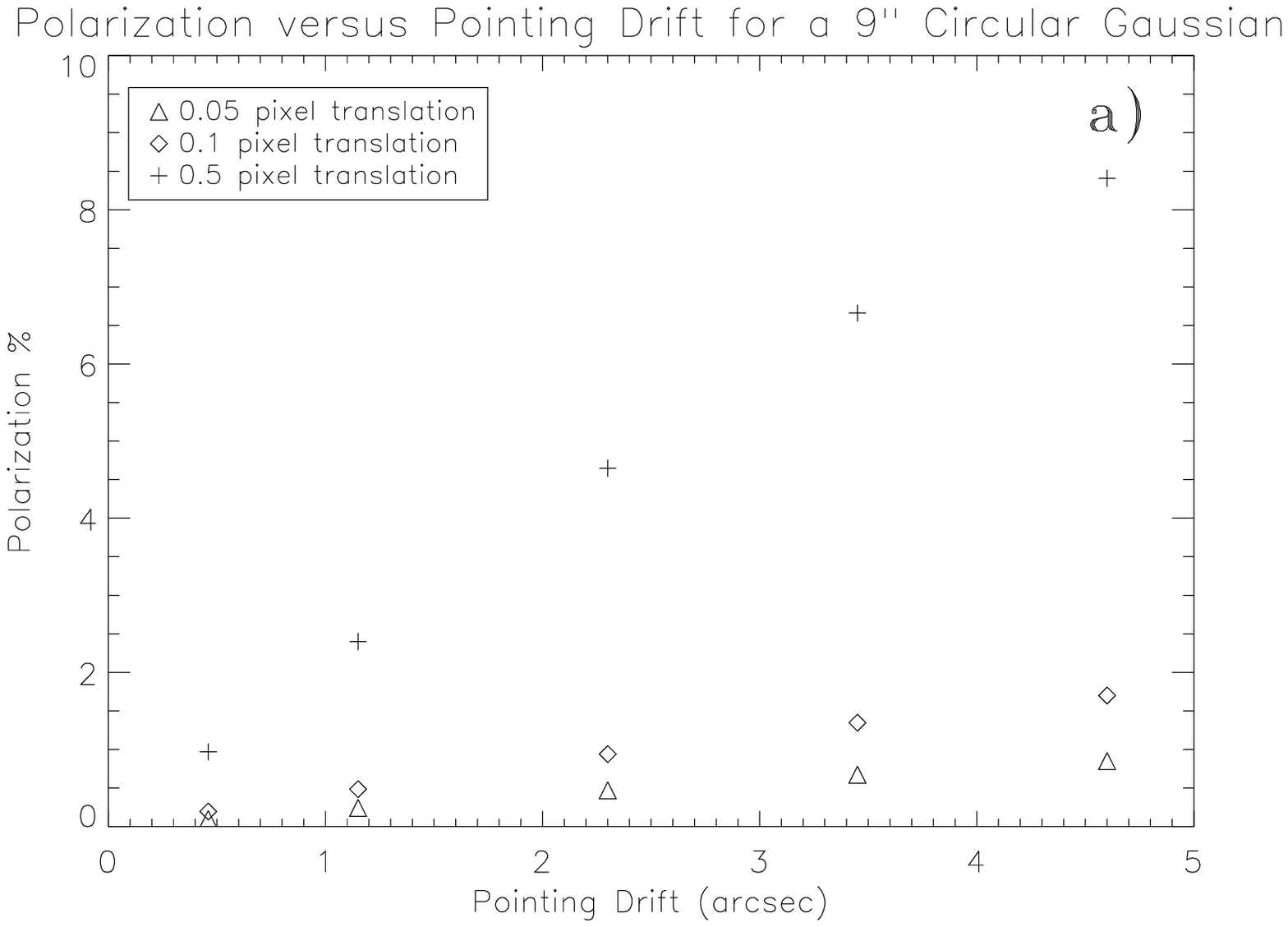}

\epsscale{0.5}
\plotone{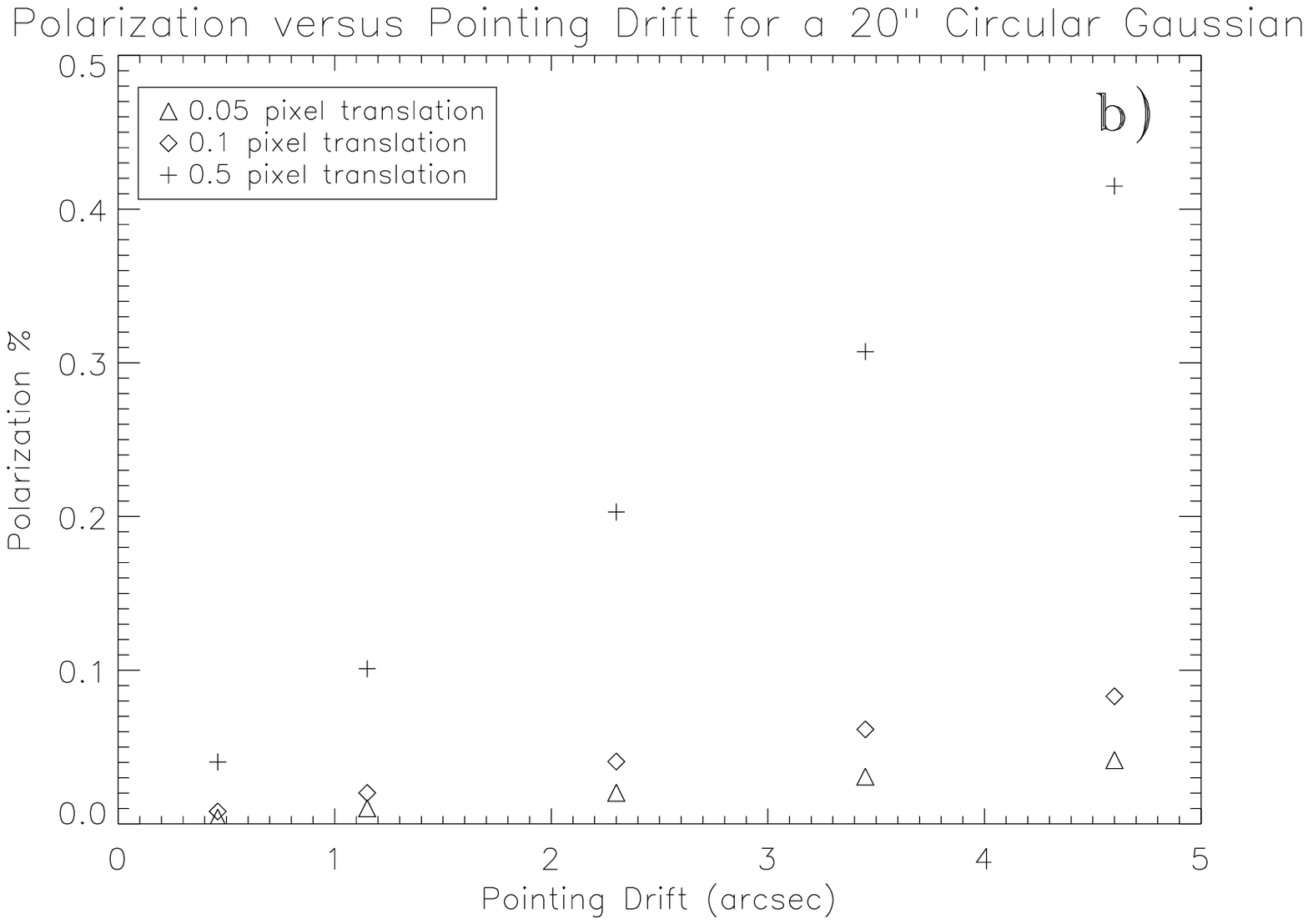}

\epsscale{0.5}
\plotone{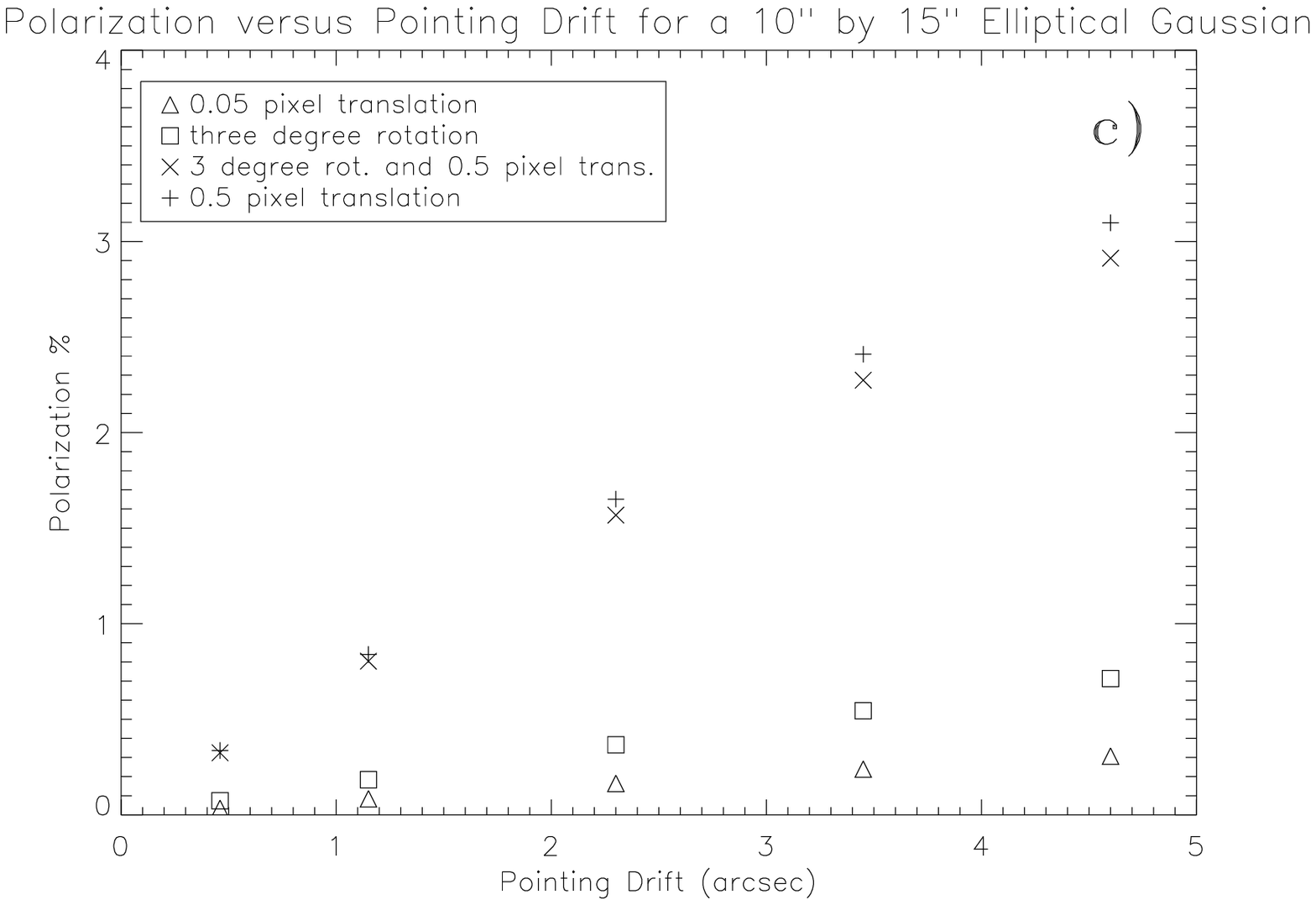}

\caption{Polarization curves as a function of pointing drifts. Each data point
represents an entire data cycle (four HWP positions). Note that only
data from the central 8 pixel $\times$ 8 pixel portion of the sub-array
was used for the analysis. Three sources were generated: a) a 9\arcsec
circular, b) a 20\arcsec circular, and c) a 10\arcsec $\times$ 15\arcsec
elliptical Gaussians, respectively. Note the various scales on the vertical axis; an indicator of the dependance of polarization percentage on source broadness.}

\end{figure}

\begin{figure}
\epsscale{0.7}
\plotone{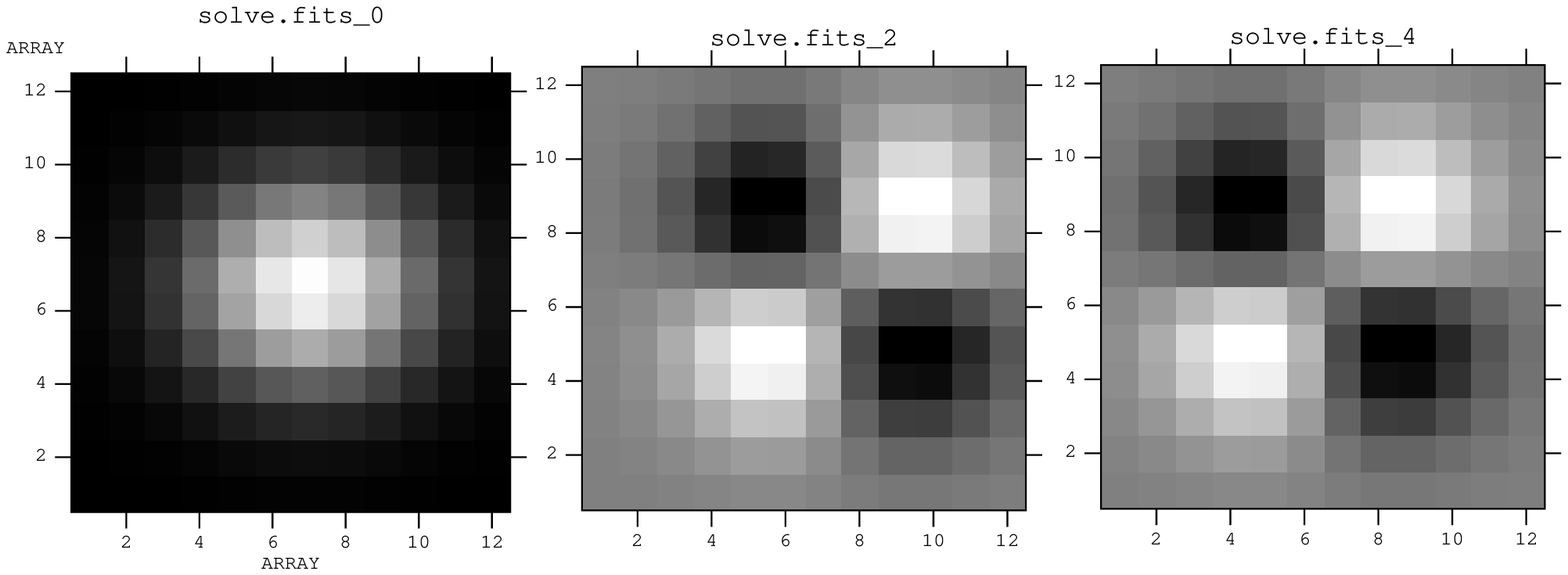}

\epsscale{0.7}
\plotone{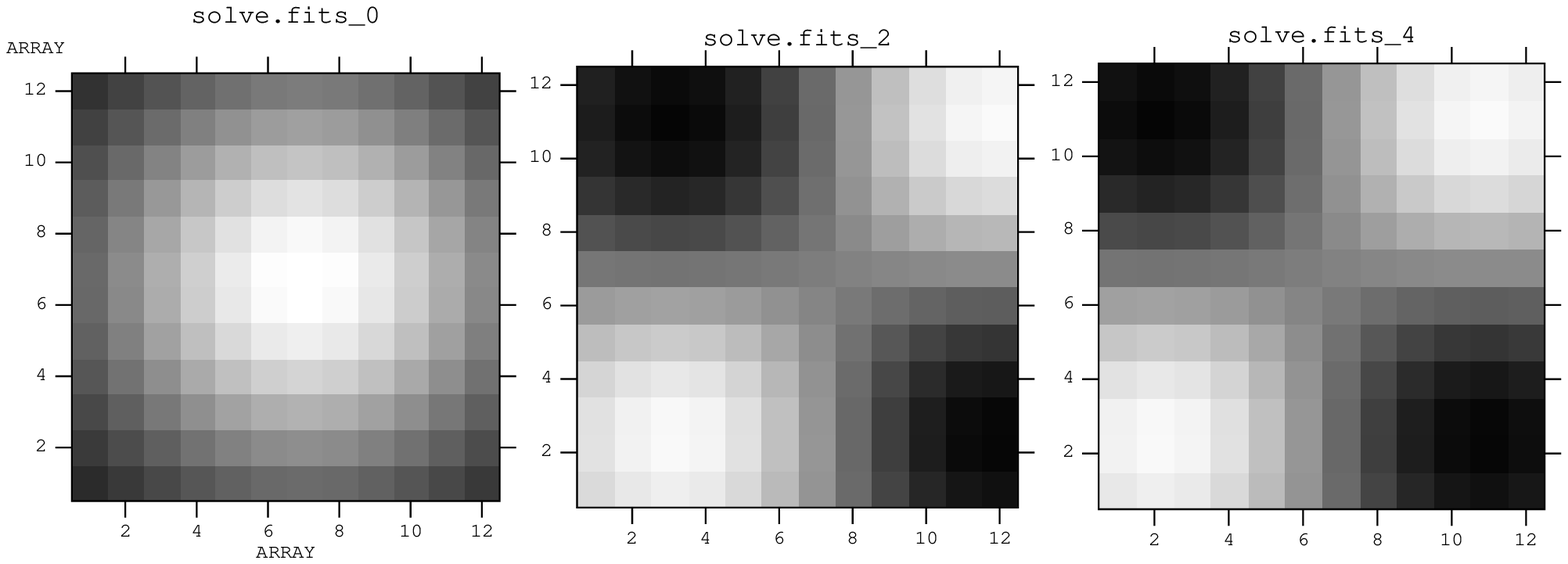}

\epsscale{0.7}
\plotone{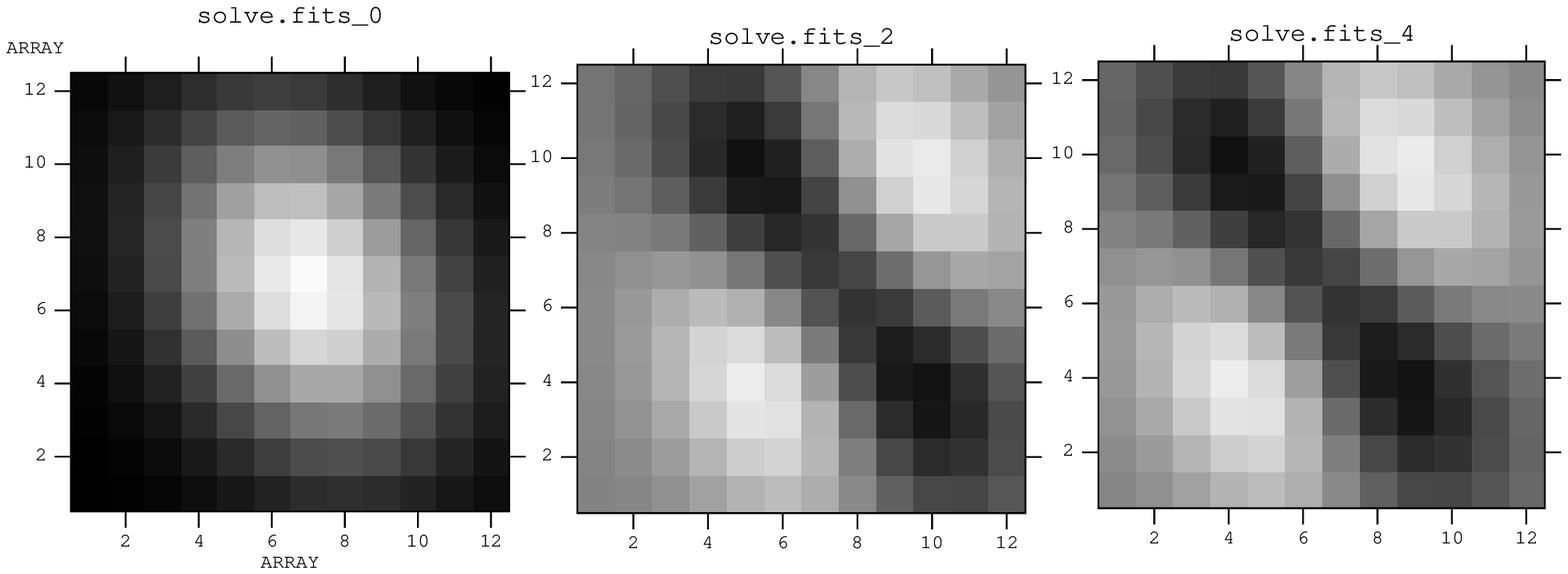}

\caption{The $I$, $Q$, and $U$ maps (from left to right) for the 9\arcsec
circular (top row), 20\arcsec circular (middle row), and 10\arcsec
$\times$ 15\arcsec elliptical (bottom row) Gaussian sources. To generate the images presented here a pointing drift of 4.6 arcseconds per HWP position (in the horizontal direction) and a translational misalignment between the H and V sub-arrays of 0.5 pixels (in the vertical direction) were applied to the simulations.  Remember that one SHARP pixel length is equivalent to 4.6 arcseconds. For the 9\arcsec
circular source, the $I$ map grey levels are at a linear scale of
0 to 1.7 (from black to white) arbitrary data units, while the $Q$
and $U$ maps are at a linear scale of -0.04 to 0.04 (from black to
white) data units. For the 20\arcsec circular source, the $I$ map
grey levels are at a linear scale of 0 to 1.9 (from black to white)
data units, while the $Q$ and $U$ maps are at a linear scale of
-0.01 to 0.01 (from black to white) data units. For the 10\arcsec
$\times$ 15\arcsec elliptical source, the $I$ map grey levels are at a
linear scale of 0 to 1.8 (from black to white) data units, while
the $Q$ and $U$ maps are at a linear scale of -0.03 to 0.03 (from
black to white) data units.}

\end{figure}

\begin{figure}
\epsscale{0.5}
\plotone{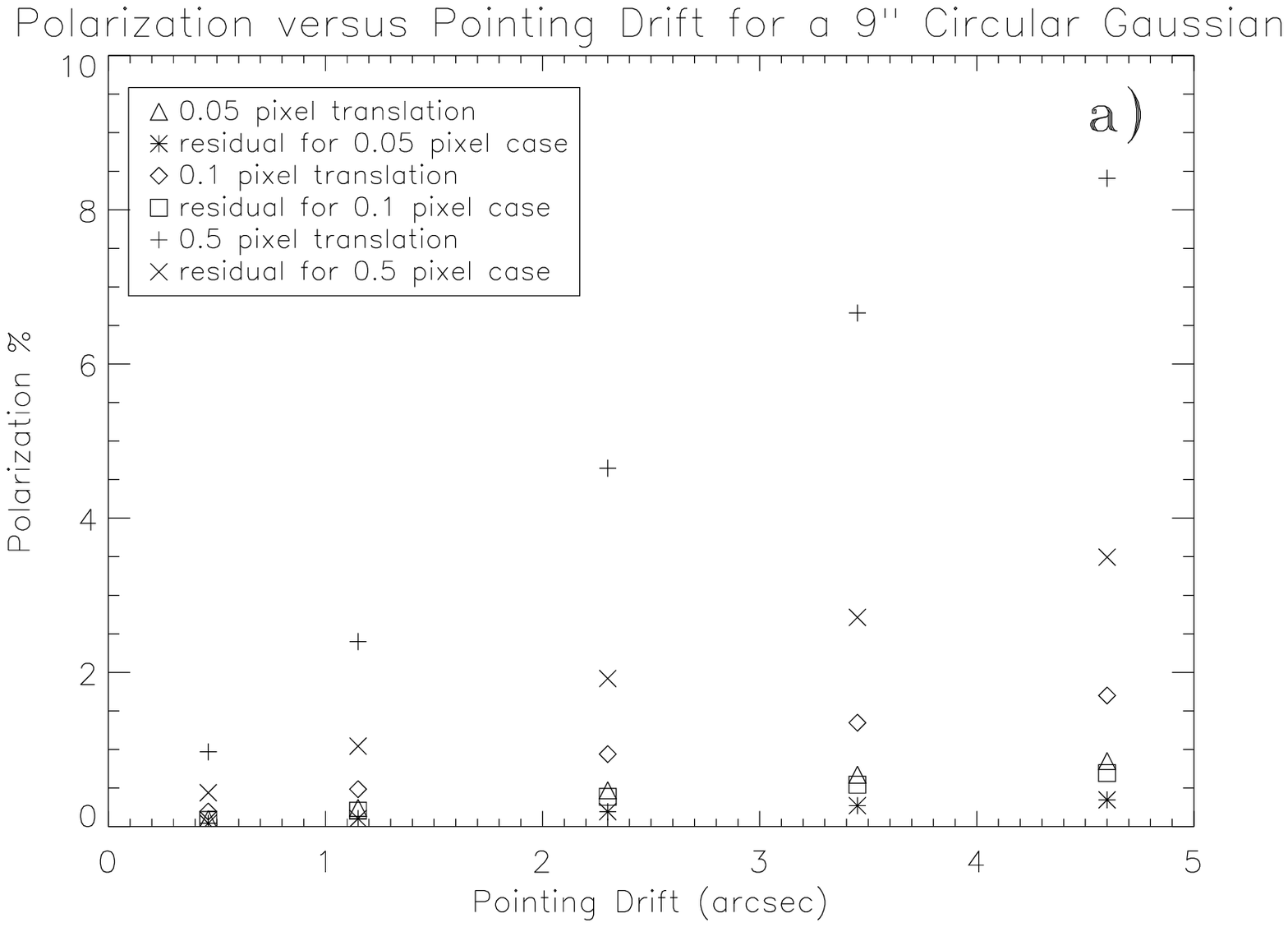}

\epsscale{0.5}
\plotone{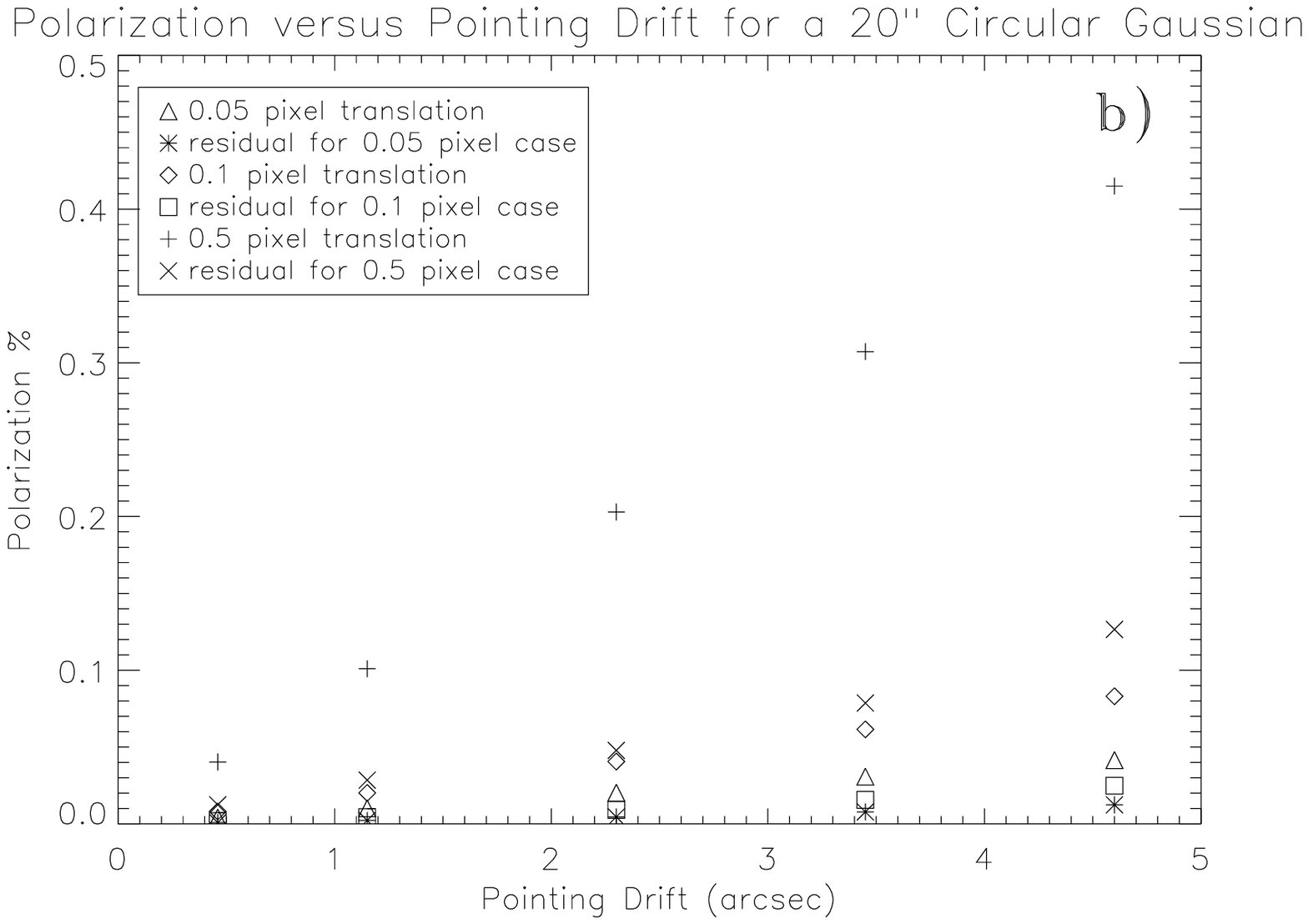}

\epsscale{0.5}
\plotone{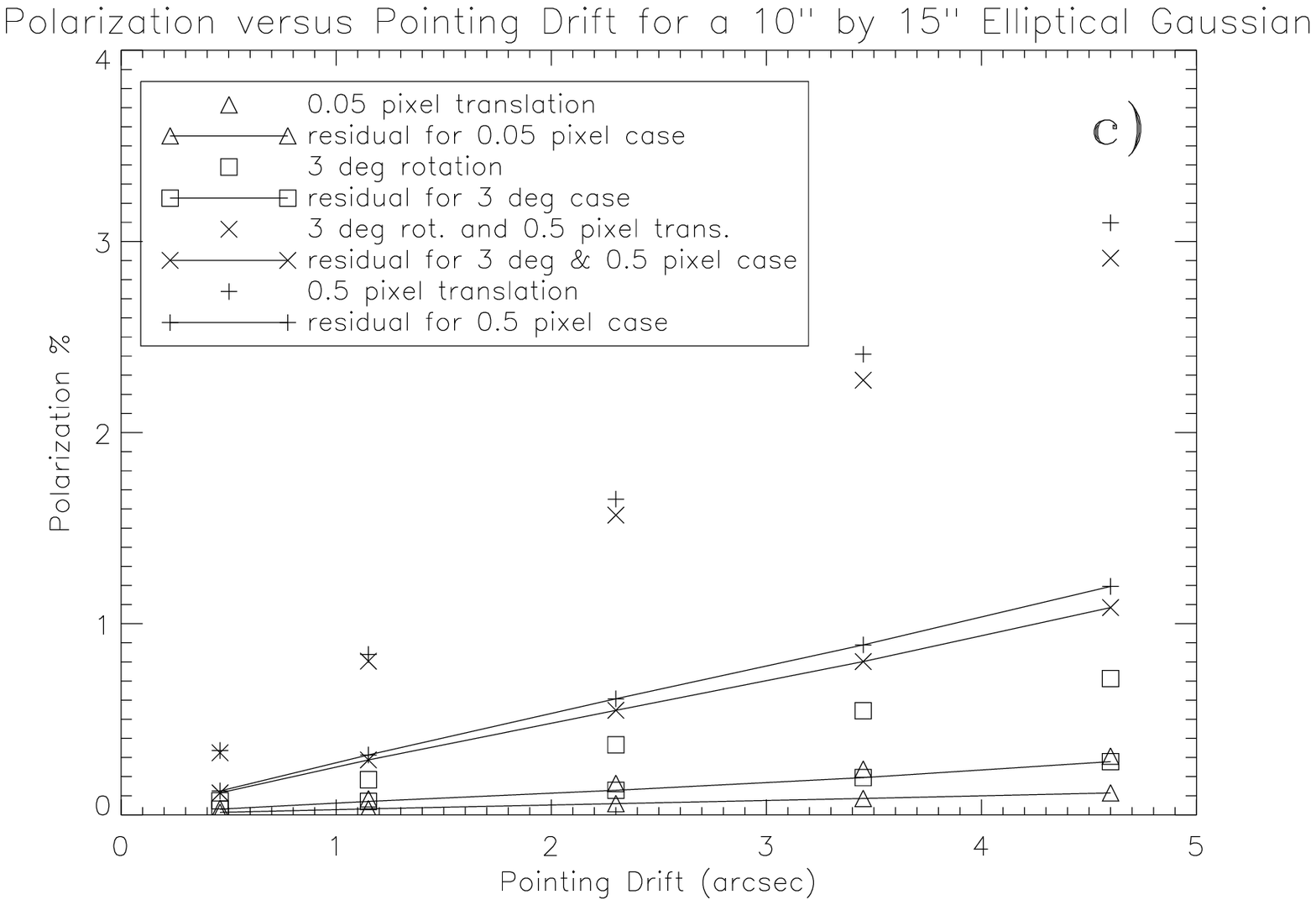}

\caption{Polarization curves as a function of pointing drifts. These plots
are identical to the ones presented in Figure 2, with the exception
that the residual polarization remaining after the correction is also
shown. }

\end{figure}

\begin{figure}
\epsscale{0.8}
\plotone{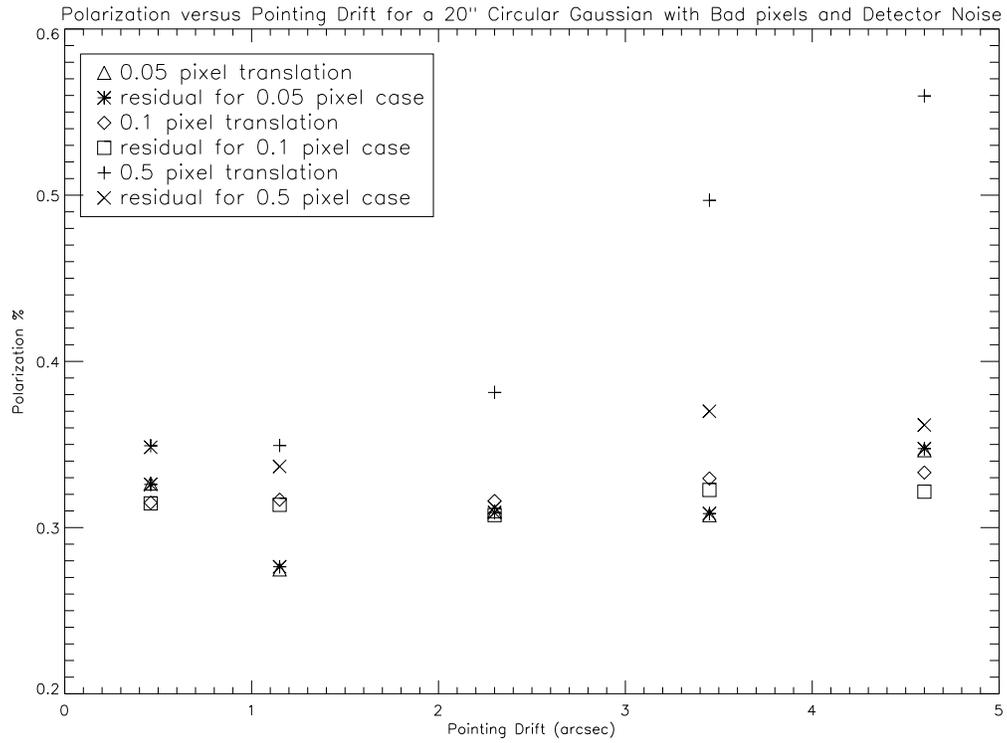}

\caption{Polarization as a function of pointing drift and translational misalignment
for the 20\arcsec circular Gaussian with bad pixels and noise introduced
into the simulation. Shown here are the induced artificial polarization
and the residual polarization after correction. }

\end{figure}

\begin{figure}
\epsscale{0.8}
\plotone{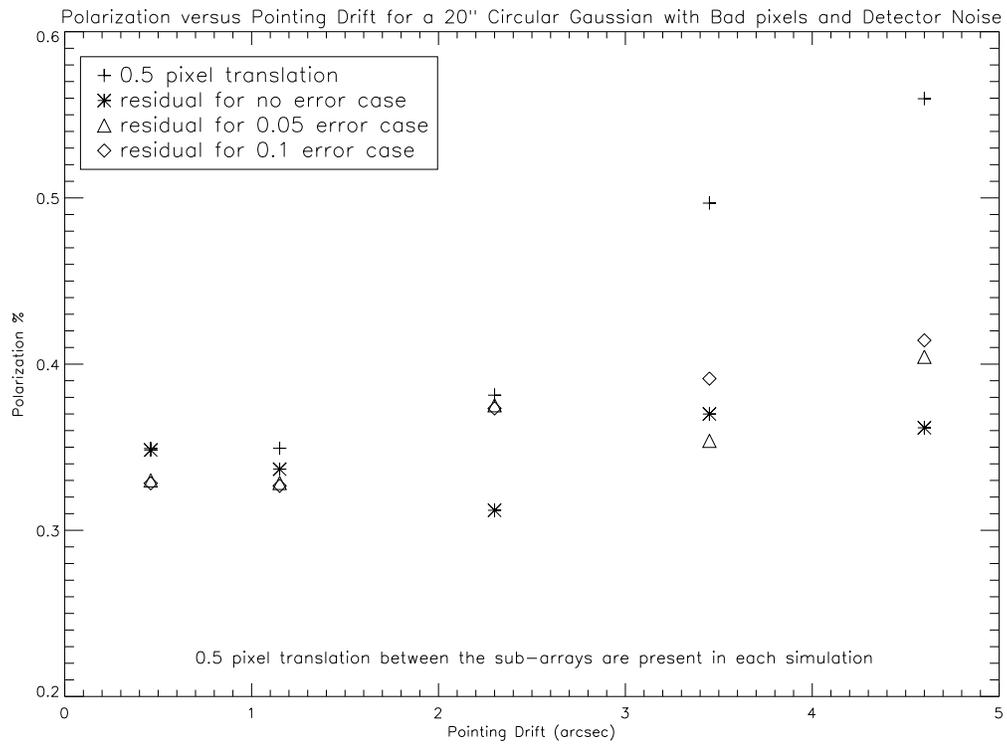}

\caption{Polarization level versus pointing drift for the 20\arcsec circular
Gaussian. Bad pixels, detector noise, and inaccuracies in the hardware
parameters are present in the analysis. }

\end{figure}

\begin{figure}
\epsscale{0.7}
\plotone{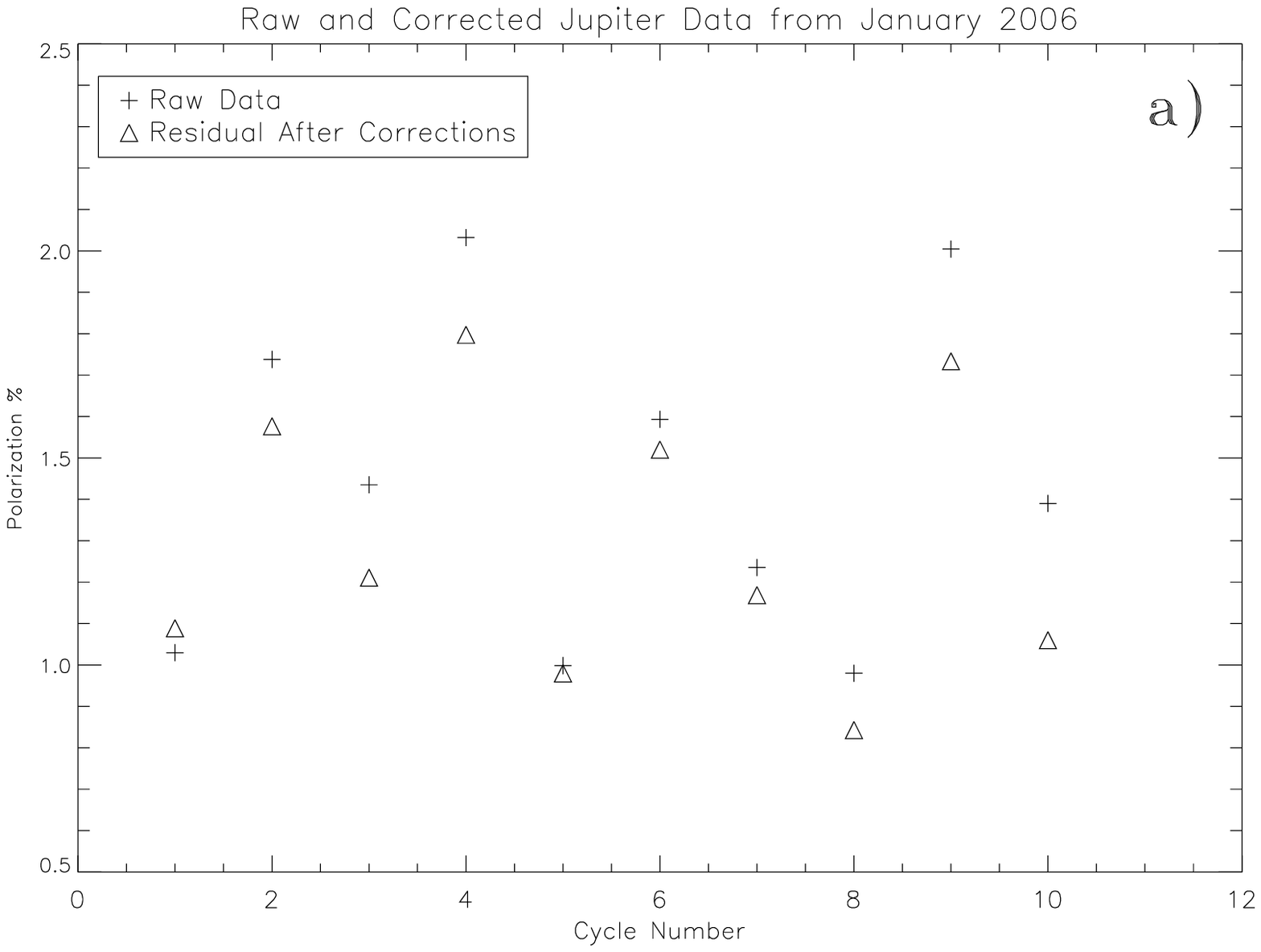}

\epsscale{0.7}
\plotone{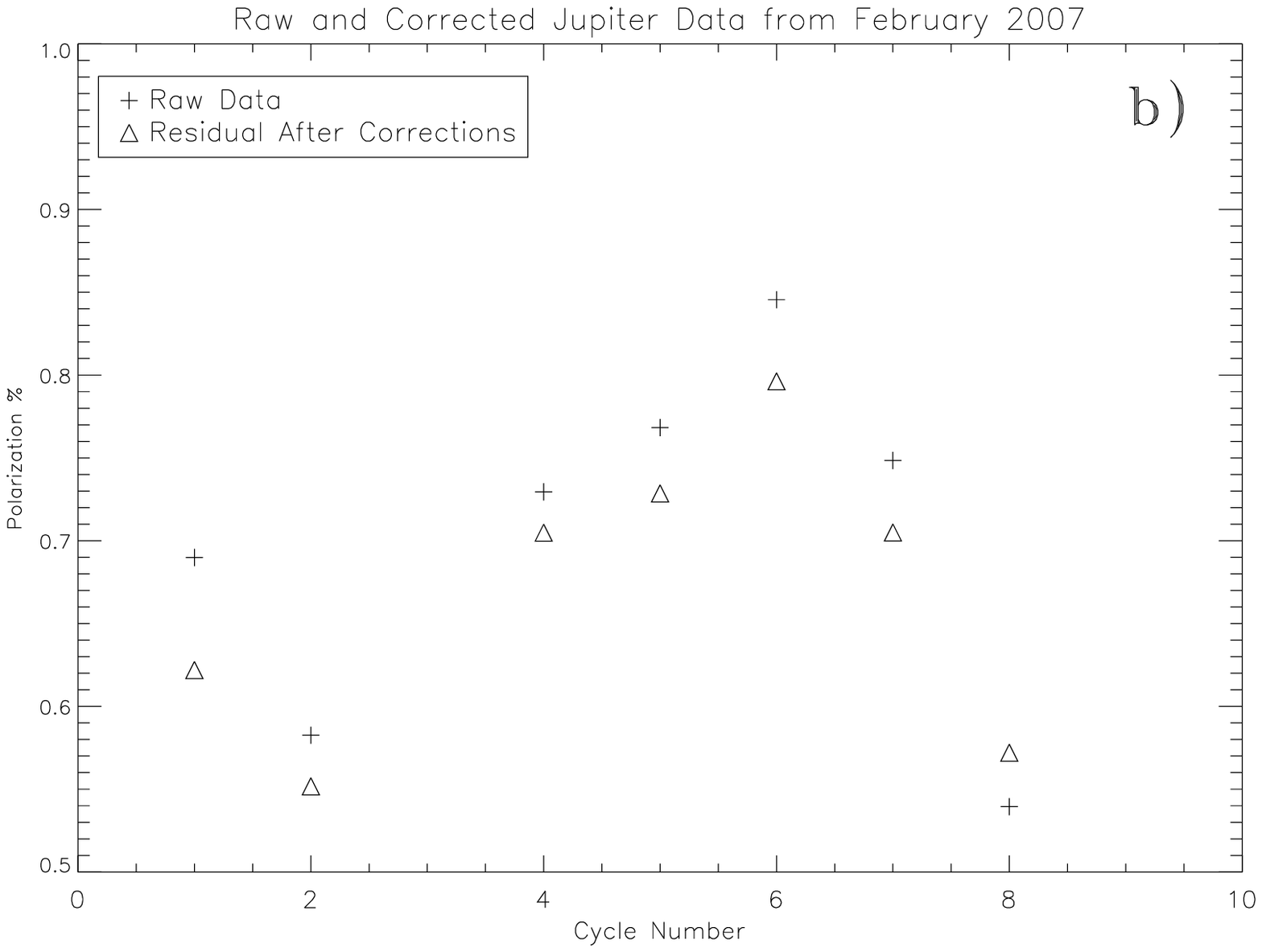}

\caption{Polarization levels before and after corrections for the artificial
polarization. Note that each cycle number refers to one HWP modulation cycles worth of data. Like the simulation analysis, only data from the central
8 pixel $\times$ 8 pixel region of the array is analyzed. Note that one
outlier is not shown at the 3rd cycle number in b), with a polarization
level of 2.7\%.}

\end{figure}

\end{document}